\documentclass{main}

\usepackage{amsmath}
\usepackage{amssymb}

\usepackage{textcomp}
\usepackage{fancyhdr}
\usepackage{lastpage}
\def\be{\begin{equation}}
\def\ee{\end{equation}}
\def\bea{\begin{eqnarray}}
\def\eea{\end{eqnarray}}

\newcommand{\Photo}{}

\fancypagestyle{firstpagefooter}{%
  \fancyfoot[L]{  \textcopyright \footnotesize This work is an open access article under a Creative Commons Attribution 4.0 International License (https://creativecommons.org/licenses/by/4.0/).}
  \fancyfoot[C]{}  
   
}

\begin{document}

\thispagestyle{firstpagefooter}
\title{\Large Partonic structure of nuclei from UPCs}

\author{\underline{V.~GUZEY}\footnote{Speaker, email: vadim.a.guzey@jyu.fi}}

\address{University of Jyvaskyla, Department of Physics, P.O. Box 35, FI-40014 University
of Jyvaskyla, Finland and Helsinki Institute of Physics, P.O. Box 64, FI-00014 University of Helsinki, Finland
}

\maketitle

\renewcommand{\thefootnote}{\fnsymbol{footnote}}
\footnotetext[1]{The author would like to dedicate this contribution to the memory of M.B.~Zhalov, his long-term colleague and collaborator.}
\renewcommand{\thefootnote}{\arabic{footnote}}

\abstracts{
We present an overview of theoretical and phenomenological studies on the partonic structure of nuclei and small-$x$ QCD dynamics using photon-nucleus $(\gamma A$) scattering in heavy ion ultraperipheral collisions (UPCs). Focusing on nuclear shadowing, we review implications of coherent charmonium and inclusive dijet production in Pb-Pb UPCs at the LHC and discuss the potential of inelastic $\gamma A$ scattering accompanied by forward neutrons in a zero degree 
calorimeter (ZDC).
}

\footnotesize DOI: \url{https://doi.org/xx.yyyyy/nnnnnnnn}


\section{Introduction}
\label{sec:intro}

The description of hard processes with nuclei in QCD involves the partonic (quark and gluon) structure
of nuclei. In the framework of collinear factorization of perturbative QCD, it is expressed in terms of nuclear parton distribution functions (nPDFs) $f_{i/A}(x,Q^2)$, which are probabilities to find parton $i$ with the momentum fraction $x$ at the resolution
scale $\sqrt{Q^2}$ in nucleus $A$. These fundamental distributions define initial conditions (cold nuclear matter effects) in proton-nucleus ($pA$) and nucleus-nucleus ($AA$) scattering and can in principle affect an onset of the regime of high parton densities (saturation) in high energy scattering with nuclei~\cite{Frankfurt:2022jns}.

The nPDFs are non-perturbative, but universal and, hence, can be determined using the global QCD fits to the available data
on various hard processes with nuclei. These typically involve deep inelastic lepton-nucleus scattering (DIS) and Drell-Yan process in $pA$ and pion-nucleus scattering off fixed nuclear targets, inclusive pion production in deuteron-gold collisions at the BNL Relativistic Heavy Ion Collider (RHIC), and gauge boson, jet and open charm production in $pA$ collisions at the CERN Large Hadron Collider (LHC). 
An incremental improvement in determination of nPDFs is one of the goals of the heavy ion program at the LHC, 
for a recent review, see~\cite{Klasen:2023uqj}. 

It also includes photon-nucleus ($\gamma A$) scattering using heavy ion ultraperipheral collisions (UPCs), which are characterized by very large 
transverse distances $|\vec{b}|$ between the colliding ions, $|\vec{b}| \gg 2 R_A$, where $R_A$ is the nucleus radius, so that the reaction is mediated by the emission of quasi-real photons in the equivalent photon approximation~\cite{Bertulani:2005ru,Baltz:2007kq}. While the exploration of the full potential of UPCs is an active field of research, the UPC measurements have already delivered important results relevant for the nPDFs and QCD dynamics at small $x$, 
which we will briefly review in this contribution. 

The global QCD analyses~\cite{Eskola:2021nhw,Duwentaster:2022kpv,AbdulKhalek:2022fyi} have established that nPDFs are not simply given by a sum of PDFs of quasi-free nucleons $f_{i/N}(x,Q^2)$, but rather are subject to distinct nuclear modifications. In this contribution, we will focus on the region of small $x$, $x < 0.05$, where nuclear shadowing suppresses $f_{i/A}(x,Q^2)$ compared to their nucleon counterparts, $f_{i/A}(x,Q^2) < A f_{i/N}(x,Q^2)$.
While it naturally reproduces the suppression of the ratio of the nucleus-to-nucleon structure functions $F_{2A}/(AF_{2N}) < 1$ for $x < 0.05$ observed in DIS on fixed nuclear targets, the same data can also be described by alternative approaches. Two notable examples include the explanation of nuclear shadowing based on the nuclear enhancement of higher-twist corrections~\cite{Qiu:2003vd} or of the saturation scale in the color dipole model~\cite{Kowalski:2007rw}. Thus, the dynamical mechanism of nuclear shadowing and its relation to the gluon 
saturation are important open questions of high energy QCD, which can be addressed now with the help of $\gamma A$ scattering at the LHC and RHIC and in the future in electron-nucleus ($eA$) scattering at the Electron-Ion Collider (EIC) in the USA~\cite{AbdulKhalek:2021gbh}.

The data-driven extraction of nPDFs using global QCD fits is agnostic to the underlying mechanism of nuclear modifications and to a certain 
degree biased toward their assumed shape, including extrapolation deep in the small-$x$ shadowing region. 
As a viable alternative, one can consider a dynamical model of nuclear shadowing referred to as the leading twist approach (LTA)~\cite{Frankfurt:2011cs} in the literature.
It is based on the combination of the Gribov-Glauber multiple scattering model of nuclear shadowing in hadron-nucleus scattering with the QCD factorization theorems for inclusive and diffractive DIS. Thus, it allows one to express the effect of nuclear shadowing for the quark and gluon nPDFs in terms of the proton diffractive PDFs. 
The resulting nPDFs are formulated at the initial scale $Q_0^2=3-4$ GeV$^2$
and serve as an input for the subsequent $Q^2$ evolution using of the usual Dokshitzer-Gribov-Lipatov-Altarelli-Parisi (DGLAP) evolution equations.
As a consequence of its connection to diffraction, LTA naturally predicts the large gluon nuclear shadowing and is characterized
by relatively small uncertainties, which result from modeling of the interaction with $N \geq 3$ nucleons of the nuclear target and which are bounded by the ``high shadowing'' and ``low shadowing'' scenarios. 

The rest of this contribution is organized as follows. In Sec.~\ref{sec:Jpsi}, we discuss 
coherent $J/\psi$ photoproduction in Pb-Pb UPCs at the LHC and its interpretation in the framework of collinear nPDFs.
The recent results on the refined treatment of the transverse plane geometry in inclusive dijet production in Pb-Pb UPCs at the LHC are
presented in Sec.~\ref{sec:dijets}. Section~\ref{sec:zcd} outlines the novel study suggesting that complementary information on nuclear shadowing
can be obtained by studying the distribution of forward neutrons produced in inelastic $\gamma A$ scattering. Finally, we give a brief summary 
and indicate new promising directions of UPC studies, emphasizing inclusive $D^0$ production, in Sec.~\ref{sec:summary}.

\section{Coherent $J/\psi$ production in Pb-Pb UPCs at the LHC}
\label{sec:Jpsi}

Exclusive photoproduction of $J/\psi$ vector mesons gives a direct access to the gluon density of the target~\cite{Ryskin:1992ui}. It is the most studied UPC process: its measurements include proton-proton ($pp$), $pA$, and $AA$ collisions, coherent (the target nucleus is intact) and incoherent (the target nucleus breaks up) cases, and UPCs accompanied by forward neutron emission due to the electromagnetic excitation of the colliding ions in different neutron classes (0n0n, 0nXn, XnXn). The measured cross sections are presented as a function of the $J/\psi$ rapidity $y$,
the momentum transfer $t$, and the  photon-nucleon center-of-mass energy $W_{\gamma p}$.

In UPCs, both ions can serve as a source of quasi-real photons and as a target. As a result, the UPCs cross section is given by a sum of two terms corresponding to the high photon momentum $k^{+}$ and the low photon momentum $k^{-}$, where $k^{\pm}=(M_{J/\psi}/2)e^{\pm y}$ and $M_{J/\psi}$ is the mass of $J/\psi$. In the case of coherent $J/\psi$ production in symmetric Pb-Pb collisions, the corresponding UPC cross section reads
\begin{equation}
\frac{d\sigma^{Pb Pb \to Pb Pb J/\psi}}{dy}=\left[k \frac{dN_{\gamma/A}}{dk} \sigma^{\gamma Pb \to J/\psi Pb}\right]_{k=k^{+}}+\left[k \frac{dN_{\gamma/A}}{dk} \sigma^{\gamma Pb \to J/\psi Pb}\right]_{k=k^{-}} \,,
\label{eq:cs_upc}
\end{equation}
In this equation, $k (dN_{\gamma/A}/dk)$ is the photon flux produced by a relativistic charge distribution, which is readily calculated
in quantum electrodynamics and includes an additional factor suppressing the soft strong interactions for small impact parameters $|\vec{b}| < 2R_A$.

Generalizing the proton result~\cite{Ryskin:1992ui} to nuclear targets, the nuclear photoproduction cross section $\sigma^{\gamma Pb \to J/\psi Pb}$ in Eq.~(\ref{eq:cs_upc}) can be written in the following form~\cite{Guzey:2013xba,Guzey:2013qza}
\begin{equation}
\sigma^{\gamma Pb \to J/\psi Pb}(W_{\gamma p})=\frac{d\sigma^{\gamma p \to J/\psi p}(W_{\gamma p},t=0)}{dt}
\left[\frac{xg_A(x,Q^2_{\rm eff})}{A xg_p(x,Q^2_{\rm eff})} \right]^2 \int_{|t_{\rm min}|}^{\infty} dt |F_A(t)|^2 \,,
\label{eq:cs_upc2}
\end{equation}
where $d\sigma^{\gamma p \to J/\psi p}/dt$ is the cross section on the proton, which was measured at HERA and also in $pp$ and $pA$ UPCs;
$xg_A(x,Q^2_{\rm eff})/[A xg_p(x,Q^2_{\rm eff})]$ is the ratio of the nucleus and proton gluon densities evaluated~\cite{Guzey:2013qza} at $x=M_{J/\psi}^2/W_{\gamma p}^2$ and  $Q^2_{\rm eff}=3$ GeV$^2$, which is predicted by the global fits of nPDFs or LTA (see Sec.~\ref{sec:intro}); 
$F_A(t)$ is the nuclear form factor; $t_{\rm min}=-x^2 m_N^2$ is the minimal momentum transfer.
Equation~(\ref{eq:cs_upc2}) has the same theoretical accuracy as the original result for the proton target~\cite{Ryskin:1992ui}: it implies the leading logarithmic approximation of perturbative QCD, the static approximation for the charmonium wave function and, additionally, assumes the factorized $t$ dependence.

\begin{figure}[t!]
  \centerline{%
    \includegraphics[width=8cm]{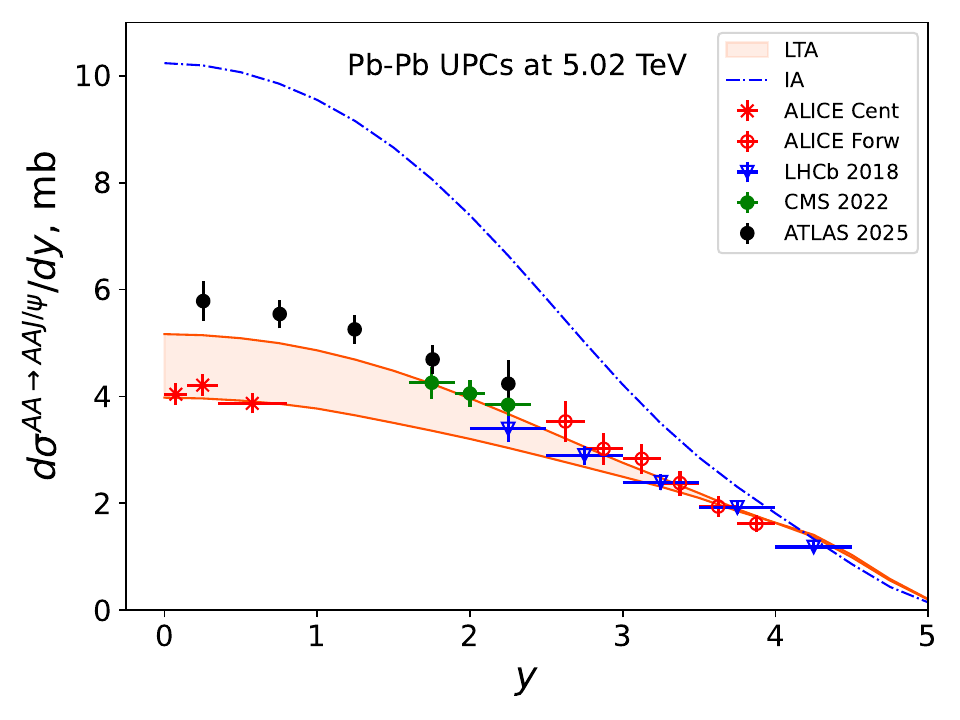}
    \includegraphics[width=8cm]{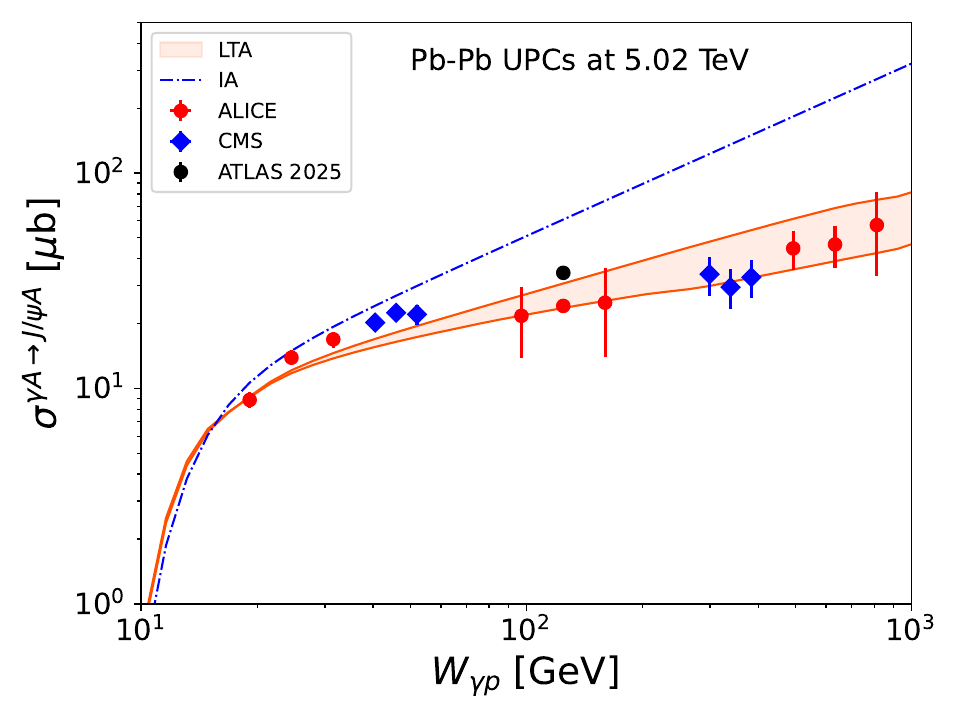}
    }
 \caption{Coherent $J/\psi$ photoproduction in Pb-Pb UPCs at 5.02 TeV: $d\sigma^{Pb Pb \to Pb Pb J/\psi}/dy$ as a function $y$ 
(left panel) and $\sigma^{\gamma Pb \to J/\psi Pb}$ as a function of $W_{\gamma p}$ (right panel).
The leading twist approach (LTA, red shaded band) and impulse approximation (IA, blue dashed line) results are compared with the 
Run 2 and Run 3 LHC data.
}
\label{fig:cs_upc}
\end{figure}

Figure~\ref{fig:cs_upc} (adapted from~\cite{Guzey:2024gff} and updated) presents the LTA predictions for $d\sigma^{Pb Pb \to Pb Pb J/\psi}/dy$ in Pb-Pb UPCs at 5.02 TeV as a function $y$ 
(left panel) and for $\sigma^{\gamma Pb \to J/\psi Pb}$ as a function of $W_{\gamma p}$ (right panel) by the red shaded band. 
Its lower and upper boundaries correspond to the ``high shadowing'' and ``low shadowing'' 
scenarios, see Sec.~\ref{sec:intro} and the detailed discussion in~\cite{Frankfurt:2011cs}.

As a reference for comparison,
the blue dashed line shows the impulse approximation (IA) result, where the photoproduction cross section is calculated neglecting nuclear modifications of the gluon density,
\begin{equation}
\sigma^{\gamma Pb \to J/\psi Pb}_{\rm IA}(W_{\gamma p})=\frac{d\sigma^{\gamma p \to J/\psi p}(W_{\gamma p},t=0)}{dt}
 \int_{|t_{\rm min}|}^{\infty} dt |F_A(t)|^2 \,.
\label{eq:cs_upc2_IA}
\end{equation}
In the left panel of Fig.~\ref{fig:cs_upc}, the theoretical predictions are compared with the ALICE~\cite{ALICE:2021gpt,ALICE:2019tqa}, LHCb~\cite{LHCb:2022ahs}, CMS~\cite{CMS:2023snh}, and the very recent Run 3 ATLAS~\cite{ATLAS:2025aav} data. In the right panel, they are compared with the CMS~\cite{CMS:2023snh}
and ALICE~\cite{ALICE:2023jgu} data as well as with the value, which we determined by extrapolating the ATLAS data~\cite{ATLAS:2025aav} to $y=0$
corresponding to $W_{\gamma p}=125$ GeV.

The most important qualitative feature of the comparison presented in Fig.~\ref{fig:cs_upc} is that the UPC data at central rapidities and large $W_{\gamma p}$ requires a large, of the order of $50-60$\%, suppression of the nuclear cross section compared to its IA estimate. As we explained in Sec.~\ref{sec:intro}, it is naturally realized in LTA and modern fits of nPDFs through the large gluon nuclear shadowing at small $x$ and
$Q^2={\cal O}(m_c^2)$, where $m_c$ is the charm quark mass.

One can see from Fig.~\ref{fig:cs_upc} that the LTA predictions, which were made more than 10 years ago, describe the Run 2 LHC data on 
$d\sigma^{Pb Pb \to Pb Pb J/\psi}/dy$ and $\sigma^{\gamma Pb \to J/\psi Pb}(W_{\gamma p})$ in the entire range of $y$ covered by the measurements (left panel)
and for $W_{\gamma p} > 100$ GeV (right panel). At the same time, the Run 3 ATLAS data, which are in tension with the ALICE Run 2 results for the central rapidity region, are somewhat underestimated. Note that the use of the state-of-the-art nPDFs~\cite{Eskola:2021nhw,Duwentaster:2022kpv,AbdulKhalek:2022fyi}  in Eq.~(\ref{eq:cs_upc2}) also leads to a good description of the presented data within the large theoretical uncertainty bands, see also Fig.~\ref{fig:S_Pb} below.

It is worth mentioning that the LTA description of the data for forward rapidities $|y| > 3$ and low energies
$W_{\gamma p} < 30$ GeV can be slightly improved by modifying the modeling of the antishadowing region $x > 0.01$, see details and examples in~\cite{Guzey:2024gff}.

It is important to contrast the description of coherent $J/\psi$ photoproduction in Pb-Pb UPCs in the framework of collinear nPDFs with that 
in the color dipole model. Qualitatively, since $J/\psi$ selects quark-antiquark dipoles of small transverse sizes corresponding to small dipole-nucleon cross sections, 
the nuclear attenuation (shadowing) effect caused by multiple interactions of these dipoles with nucleons of the nuclear target 
is not expected to be sizable~\cite{Frankfurt:2002kd}. Thus, almost independently of the underlying dynamics of the dipole-nucleon interaction (the presence of gluon saturation or hot spots), the dipole model framework predicts a weaker nuclear suppression than that 
in the collinear approach. As a consequence, the dipole model somewhat overestimates the Run 2 ALICE data, but is compatible with the 
cross section measured by ATLAS~\cite{ATLAS:2025aav,Mantysaari:2022sux,Mantysaari:2023xcu}.

It is convenient to convert the UPC cross section~(\ref{eq:cs_upc}) and the photoproduction cross section~(\ref{eq:cs_upc2})
into the nuclear suppression factor $S_{Pb}$, which can be defined as follows~\cite{Guzey:2013xba,Guzey:2013qza}
\begin{equation}
S_{Pb}(x)=\left[\frac{\sigma^{\gamma Pb \to J/\psi Pb}(W_{\gamma p})}{\sigma^{\gamma Pb \to J/\psi Pb}_{\rm IA}(W_{\gamma p})} \right]^{1/2}
= \frac{xg_A(x,Q^2_{\rm eff})}{A xg_p(x,Q^2_{\rm eff})} \,,
\label{eq:S_Pb}
\end{equation}
where $x=M_{J/\psi}^2/W_{\gamma p}^2$. The last equality is readily obtained using Eqs.~(\ref{eq:cs_upc2}) and (\ref{eq:cs_upc2_IA}).
The advantages of introducing $S_{Pb}$ are that it ``resolves'' the two-fold photon energy ambiguity in Eq.~(\ref{eq:cs_upc}) 
and allows for a direct comparison with the $g_A/(A g_p)$ ratio of the nuclear and proton gluon densities.

The left panel of Fig.~\ref{fig:S_Pb} (adapted from~\cite{Guzey:2024gff} and updated) shows the nuclear suppression factor $S_{Pb}$ as a function
of $x$ and compares the CMS~\cite{CMS:2023snh} and ALICE~\cite{ALICE:2023jgu} experimental values for it, as well as the ones that we extracted from the 
ATLAS~\cite{ATLAS:2025aav} and STAR~\cite{STAR:2023vvb} data, with the theoretical predictions for $g_A/(A g_p)$ at $Q^2_{\rm eff}=3$ GeV$^2$ obtained using LTA and 
the EPPS21~\cite{Eskola:2021nhw}, nCTEQ15HQ~\cite{Duwentaster:2022kpv}, and nNNPDF3.0~\cite{AbdulKhalek:2022fyi} nPDFs. One can see from the figure that LTA provides a good description of the data for $x < 10^{-3}$ (with the exception of the 2025 ATLAS data point), which gives the direct evidence for the large gluon nuclear shadowing: $R_g=g_A/(A g_p) \approx 0.6$ at $x= 6 \times 10^{-4}-10^{-3}$, with an indication of further decreasing $R_g$ down to $x\approx 10^{-5}$.
At the same time, LTA underestimates $S_{Pb}(x)$ around $x=0.01$ because of the non-negligible gluon nuclear shadowing.

One can also see from the left panel of Fig.~\ref{fig:S_Pb} that within significant theoretical uncertainties, the modern nuclear PDFs (EPPS21, nCTEQ15HQ, nNNPDF3.0) result in the values for $S_{Pb}(x)$, see Eq.~(\ref{eq:S_Pb}), which agree 
with the data, and predict an essentially flat behavior of $S_{Pb}(x)$ for $x < 10^{-3}$.

\begin{figure}[t!]
  \centerline{%
    \includegraphics[width=8cm]{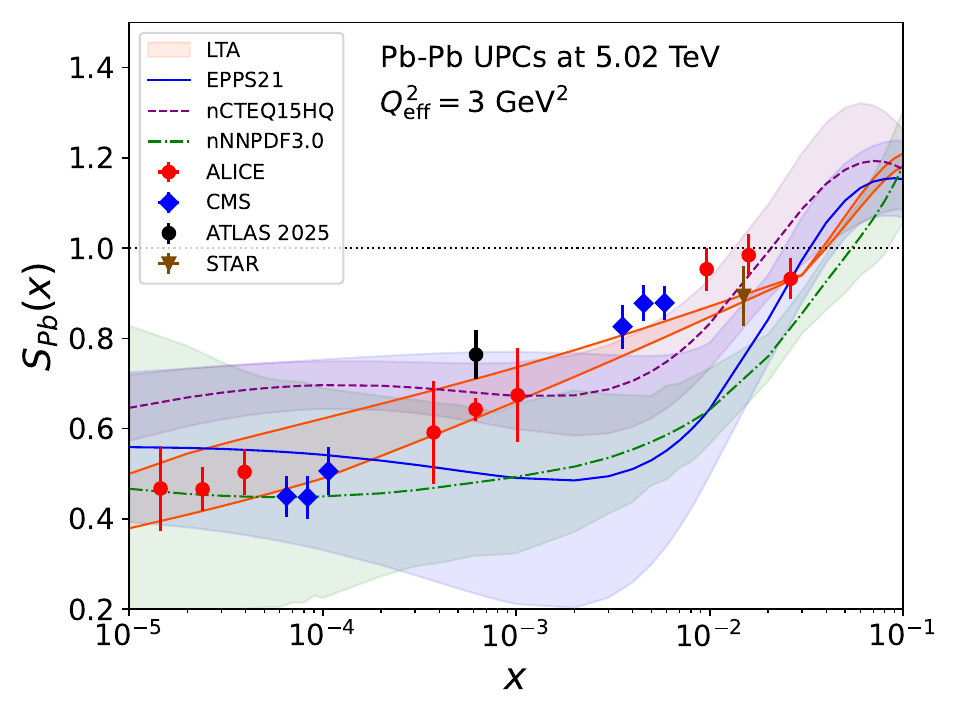}
    \includegraphics[width=8cm]{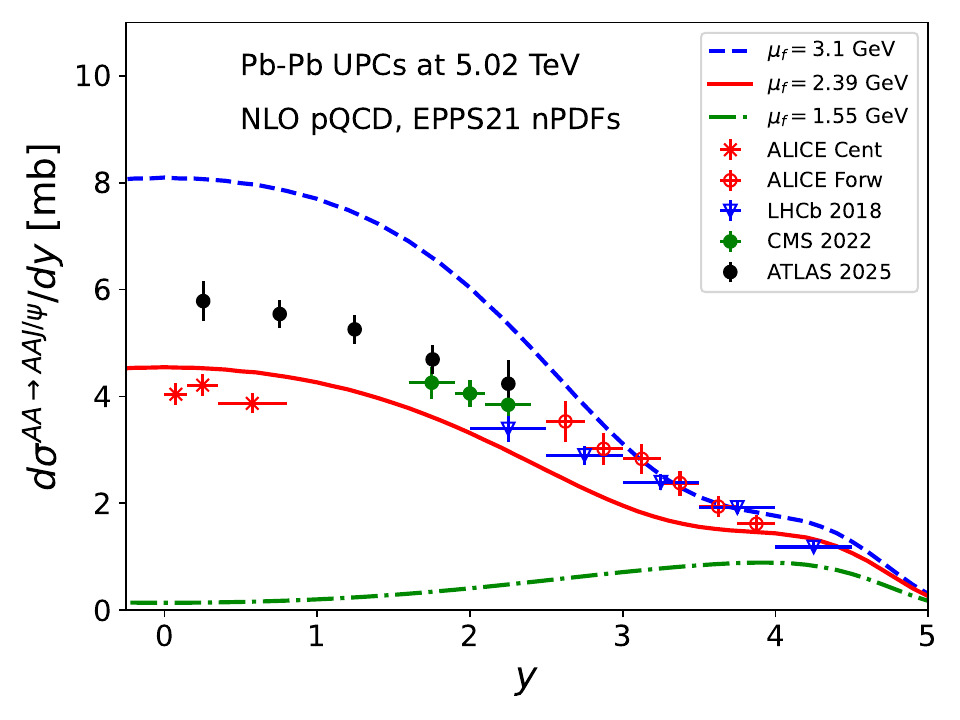}
    }
 \caption{(Left) The nuclear suppression factor $S_{Pb}$ as a function of $x$: The CMS and ALICE Run 2 data along with the values extracted from
 the Run 3 ATLAS and STAR data and their comparison with the theoretical predictions for $g_A/(A g_p)$ using the LTA and 
 modern (EPPS21, nCTEQ15HQ, nNNPDF3.0) nuclear PDFs. 
 (Right) The $d\sigma^{Pb Pb \to Pb Pb J/\psi}/dy$ cross section of coherent $J/\psi$ photoproduction in Pb-Pb UPCs at 5.02 TeV: the NLO pQCD results at the three values of $\mu_f$ vs.~the Run 2 and Run 3 LHC data.
}
\label{fig:S_Pb}
\end{figure}

Going beyond the leading logarithmic approximation of Eq.~(\ref{eq:cs_upc2}), one can show that to the next-to-leading (NLO) accuracy of perturbative QCD, the amplitude of exclusive $J/\psi$ photoproduction can be expressed as convolution of the gluon and quark hard coefficient functions $T_{g,q}$ with the gluon and quark generalized parton distributions (GPDs) of the target~\cite{Ivanov:2004vd}. In this framework, the $\gamma A \to J/\psi A$ amplitude of coherent 
$J/\psi$ photoproduction on a nuclear target reads~\cite{Eskola:2022vpi,Eskola:2022vaf}
\begin{equation}
{\cal M}^{\gamma A \to J/\psi A}(t) \propto \sqrt{\langle O_1 \rangle_{J/\psi}} \int^{1}_{-1} dx \left[T_g(x,\xi,m_c/\mu_f) F_A^g(x,\xi,t,\mu_f)+T_q(x,\xi,m_c/\mu_f) F_A^q(x,\xi,t,\mu_f) \right] \,,
\label{eq:amp_nlo}
\end{equation}
where $\langle O_1 \rangle_{J/\psi}$ is the matrix element describing the $J/\psi \to l^{+} l^{-}$ leptonic decay rate in nonrelativistic QCD;
$F_A^{g,q}$ are the gluon and quark nuclear GPDs; $m_c=M_{J/\psi}/2$ and $\mu_f$ are the charm quark mass and the factorization scale, respectively; 
$\xi \approx x/2= M_{J/\psi}^2/(2 W_{\gamma p}^2) \ll 1$ is the skewness longitudinal momentum fraction transferred to the target.
Note that unlike the leading order (LO) expression of Eq.~(\ref{eq:cs_upc2}), both gluons and quarks contribute at NLO.

While GPDs in general differ from PDFs, one can reliably relate the two distributions at small $\xi$ and use the following forward model for nuclear 
GPDs,
\begin{eqnarray}
F_A^{g}(x,\xi,t,\mu_f) &=& xg_A(x,\mu_f) F_A(t) \,, \nonumber\\
F_A^{q}(x,\xi,t,\mu_f) &=& q_A(x,\mu_f) F_A(t) \,,
\label{eq:gpd}
\end{eqnarray}
where $g_A$ and $q_A$ are the usual gluon and quark nPDFs, and $F_A(t)$ is the nuclear form factor normalized to unity, $F_A(t=0)=1$.
Note that similarly to Eq.~(\ref{eq:cs_upc2}), the conjecture of Eq.~(\ref{eq:gpd}) implies the factorized $t$ dependence. This is sufficiently
accurate for the $t$-integrated $\gamma A \to J/\psi A$ cross section, but is clearly inadequate for the differential cross section (see the discussion below). 

The right panel of Fig.~\ref{fig:S_Pb} presents the NLO pQCD results~\cite{Eskola:2022vpi,Eskola:2022vaf} for the cross section of coherent $J/\psi$ photoproduction in Pb-Pb UPCs at 5.02 TeV as a function of $y$. They are based on Eqs.~(\ref{eq:amp_nlo}) and (\ref{eq:gpd}), where the latter employs 
the central value of the EPPS21 nPDFs. The calculations were performed at the three different values of $\mu_f=(1.55, 2.39, 3.1)$ GeV and keeping
$\mu_R=\mu_f$, where $\mu_R$ is the renormalization scale. As in the left panel of Fig.~\ref{fig:cs_upc}, they are compared with the Run 2 and Run 3 LHC data.

One can see from the figure that the calculated cross section exhibits a very strong factorization scale dependence, which confirms its earlier observation~\cite{Ivanov:2004vd}. Nevertheless, one can find an ``optimal scale'', $\mu_f=2.39$ GeV in the case of the EPPS21 nPDFs, which provides 
a reasonable description of the Run 1 and 2 LHC data~\cite{Eskola:2022vpi,Eskola:2022vaf}.
Note also that the nuclear PDF uncertainties (not shown) are also very significant (they are relatively as large as twice those shown in the left panel of Fig.~\ref{fig:S_Pb}), which in principle allows one to accommodate all the data shown in the right panel of Fig.~\ref{fig:S_Pb} within the NLO pQCD
approach~\cite{Eskola:2022vpi,Eskola:2022vaf}.

It is well understood that the very strong $\mu_f$ dependence is caused by large cancellations between the LO and NLO gluon contributions because the 
latter is enhanced at small $\xi$. This increases the role of the light quark contribution to the $\gamma A \to J/\psi A$ process, challenges the interpretation of the UPC data in terms of the nuclear gluon density, and in general raises the question of stability of
the perturbation series for this process. A possible solution to tame this effect was offered in~\cite{Flett:2024htj}, 
where it was shown that using the formalism of high-energy factorization (HEF), one can resum higher-order, $\ln(1/\xi)$-enhanced QCD corrections to the gluon and quark NLO coefficient functions. This reduces the dependence on $\mu_f$ and especially on $\mu_R$ and largely restores the gluon dominance in the 
considered process.

Another complication, which may affect the theoretical interpretation of the data on coherent $J/\psi$ photoproduction in UPCs, is related to potentially
significant relativistic corrections to the charmonium wave function. While they are usually assumed to be small compared to
the scale uncertainties in the collinear approach, the analyses within the dipole model framework found them to be phenomenologically important~\cite{Frankfurt:1997fj,Lappi:2020ufv,Mantysaari:2025idf}.

In the case of exclusive $\Upsilon$ photoproduction, many theoretical issues mentioned above are much milder: 
the NLO corrections to the pQCD cross section are moderate~\cite{Ivanov:2004vd}, 
the relativistic corrections to the bottomonium wave function are expected to be small, and the modeling of GPDs benefits from the longer $Q^2$ evolution of GPDs~\cite{Dutrieux:2023qnz} up to $\mu_f={\cal O}(m_b)$, where $m_b$ is the mass of the bottom quark. 
 
The first NLO pQCD predictions for the cross section of coherent  $\Upsilon$ photoproduction in Pb-Pb UPCs at 5.02 TeV as a function of the rapidity $y$,
including GPD modeling through the Shuvaev integral transform, 
were made in~\cite{Eskola:2023oos}. It was explicitly shown that the GPD effects are small, i.e., one can reliably use Eq.~(\ref{eq:gpd}), 
the scale uncertainties are significantly reduced compared to the $J/\psi$ case, and the cross section is largely dominated by the gluon contribution.
These predictions probe the scale $Q^2$ dependence of nuclear modifications of the nuclear GPDs and PDFs and 
broadly agree with the recent CMS measurement of this process~\cite{CMS:2025bsz}.

Important information on the small-$x$ QCD dynamics of nuclear shadowing comes from the dependence of the differential nuclear photoproduction cross
section $d\sigma^{\gamma Pb \to J/\psi Pb}/dt$ on the momentum transfer $t$. 
Qualitatively, the LTA model predicts that the suppression of nPDFs due to nuclear shadowing is stronger at the nucleus center than at its 
periphery, which leads to the broadening of nPDFs in the impact parameter space~\cite{Frankfurt:2011cs}. It directly affects
the $t$ dependence of $d\sigma^{\gamma Pb \to J/\psi Pb}/dt$ and causes a shift of the diffractive dip toward the smaller values of $|t|$ 
compared to the IA estimate~\cite{Guzey:2016qwo}. This prediction was convincingly confirmed by the ALICE measurement~\cite{ALICE:2021tyx}.

Expressing the shift of the diffractive dip in terms of the effective nuclear gluon radius $R_A^{(g)}$, one finds~\cite{Guzey:2016qwo} 
that the leading twist nuclear shadowing increases it. For instance, at $x=10^{-3}$ and $Q^2_{\rm eff}=3$ GeV$^2$, 
\begin{equation}
\frac{\Delta R_A^{(g)}}{R_A^{(g)}}= 1.05 - 1.11 \,,
\label{eq:R_A}
\end{equation}
where the lower and upper limits corresponds to the ``low shadowing'' and ``high shadowing'' scenarios.
Since the diffractive dip position and $\Delta R_A^{(g)}/R_A^{(g)}$ directly depend on the magnitude of the gluon nuclear shadowing, they can in principle 
be used to refine parameters of the models of nuclear shadowing and possibly even to discriminate among them. However, 
the ALICE result~\cite{ALICE:2021tyx} does not appear to have a strong discriminating power and is also consistent with the similar effect predicted in the dipole model framework with gluon saturation~\cite{Mantysaari:2022sux,Bendova:2020hbb}.

\section{Inclusive dijet photoproduction in Pb-Pb UPCs}
\label{sec:dijets}

Production of jets in $pA$ scattering at the LHC has played an important role in constraining the partonic structure of nuclei, 
in particular the nuclear gluon distribution in heavy nuclei~\cite{Eskola:2021nhw}. Similarly, photoproduction of dijets in UPCs provides additional 
constraints on nPDFs at intermediate $x$ and large $Q^2$, which are complementary to those at very small $x$ and $Q^2={\cal O}(m_c^2,m_b^2)$ in exclusive production of 
quarkonia discussed in Sec.~\ref{sec:Jpsi}. In particular, the very first measurement of photonuclear jet production in Pb-Pb UPCs at 5.02 TeV by ATLAS~\cite{ATLAS:2024mvt} has covered the region of $0.002 < x < 0.5$ and $35 < \sqrt{Q^2} < 212.5$ GeV (here we identified the hard scale 
with the total transverse momentum of the jets $H_T$).

The NLO pQCD predictions for inclusive and diffractive dijet photoproduction in Pb-Pb UPCs in the LHC kinematics are summarized in~\cite{Guzey:2024jvi}.
Below we focus on the more recent calculation~\cite{Eskola:2024fhf}, which involves a refined analysis of the transverse plane geometry 
and takes into account the experimental condition requiring zero neutrons on the photon-emitting side and at least one neutron on the nuclear target side 
(the 0nXn neutron class).

In the framework of collinear factorization of pQCD, the cross section of inclusive dijet photoproduction in
heavy ion UPCs, $AA \to A+2{\rm jets}+X$, can be written as convolution of the photon flux $f_{\gamma/A}$, the photon PDFs
$f_{a/\gamma}$ (for the resolved photon contribution), the nuclear PDFs $f_{b/A}$, and the partonic cross section 
$d\hat{\sigma}^{ab \to {\rm jets}}$,
\begin{eqnarray}
d\sigma^{AA \to A+ 2{\rm jets} + X} &=& \sum_{a,b} \int dy \int dx_{\gamma} \int dx_A \left[\int d^2 \vec{b} \,\Gamma_{AA}(\vec{b}) 
\int d^2 \vec{r} \int d^2 \vec{s}\, \delta^2(\vec{r}-\vec{s}-\vec{b})\right]\nonumber\\
&\times&
f_{\gamma/A}(y,\vec{r}) f_{a/\gamma}(x_{\gamma},Q^2)
f_{b/A}(x_A,Q^2,\vec{s}) d\hat{\sigma}^{ab \to {\rm jets}}\,,
\label{eq:upc_dijets}
\end{eqnarray}
where the longitudinal momentum fractions $y$, $x_{\gamma}$, and $x_A$ refer to the equivalent photon, the photon PDFs, and the nuclear PDFs, respectively;
$a$ and $b$ are the parton flavors.
 In the direct photon case, one identifies $a=\gamma$ and uses
$f_{\gamma/\gamma}(x_{\gamma},Q^2)=\delta(1-x_{\gamma})$ at LO and a more involved procedure at NLO because the separation of the direct and resolved photon contributions depends on the factorization scheme and scale. 

The expression in the square brackets in Eq.~(\ref{eq:upc_dijets}) was first introduced in~\cite{Eskola:2024fhf}, and it  takes into account that the distance $\vec{r}$ between the center of the photon-emitting nucleus and transverse position $\vec{s}$ of a parton in the target nucleus is not exactly equal to the impact parameter $\vec{b}$ 
(the transverse distance between the centers of the colliding nuclei), when $|\vec{b}|$ is not too large. Neglecting this effect amounts to dropping 
$\delta^2(\vec{r}-\vec{s}-\vec{b})$ in Eq.~(\ref{eq:upc_dijets}), which after the integration over $d^2 \vec{r}$ and $d^2 \vec{s}$ leads to the commonly used expression for the dijet UPC cross section~\cite{Guzey:2024jvi,Guzey:2018dlm}.

The factor of $\Gamma_{AA}(\vec{b})$ gives the probability of not having the strong hadronic interactions at the impact parameter $\vec{b}$ between the 
colliding nuclei. It is an essential ingredient of the photon flux calculation, which allows one to refine the frequently used point-like (PL) approximation 
for it and, hence, circumvents the need to introduce the impact parameter cutoff.

The numerical analysis in~\cite{Eskola:2024fhf} has shown that with a very high accuracy, the impact-parameter dependent nuclear PDFs can be written in the following simple form 
\begin{equation}
f_{b/A}(x_A,Q^2,\vec{s})=T_A(\vec{s}) f_{b/A}(x_A,Q^2) \,,
\label{eq:nPDFs_impact}
\end{equation}
where $T_A(\vec{s})=\int dz \rho_A(\vec{s},z)$ is the nuclear thickness function normalized to unity with $\rho_A$ being the nuclear density.
Note that the approximation of Eq.~(\ref{eq:nPDFs_impact}) is the same as the one in Eq.~(\ref{eq:gpd}). Substituting Eq.~(\ref{eq:nPDFs_impact}) in Eq.~(\ref{eq:upc_dijets}), one obtains 
\begin{equation}
d\sigma^{AA \to A+ 2{\rm jets} + X} = \sum_{a,b} \int dy \int dx_{\gamma} \int dx_A
f_{\gamma/A}^{\rm eff}(y) f_{a/\gamma}(x_{\gamma},Q^2)
f_{b/A}(x_A,Q^2) d\hat{\sigma}^{ab \to {\rm jets}}\,,
\label{eq:upc_dijets2}
\end{equation}
where $f_{\gamma/A}^{\rm eff}(y)$ is the effective photon flux, 
\begin{equation}
f_{\gamma/A}^{\rm eff}(y)=\int d^2 \vec{r} \int d^2 \vec{s}\, f_{\gamma/A}(y,\vec{r}) T_A(\vec{s}) \Gamma_{AA}(\vec{r}-\vec{s}) \,.
\label{eq:flux_effective}
\end{equation}

\begin{figure}[t!]
  \centerline{%
    \includegraphics[width=8.cm]{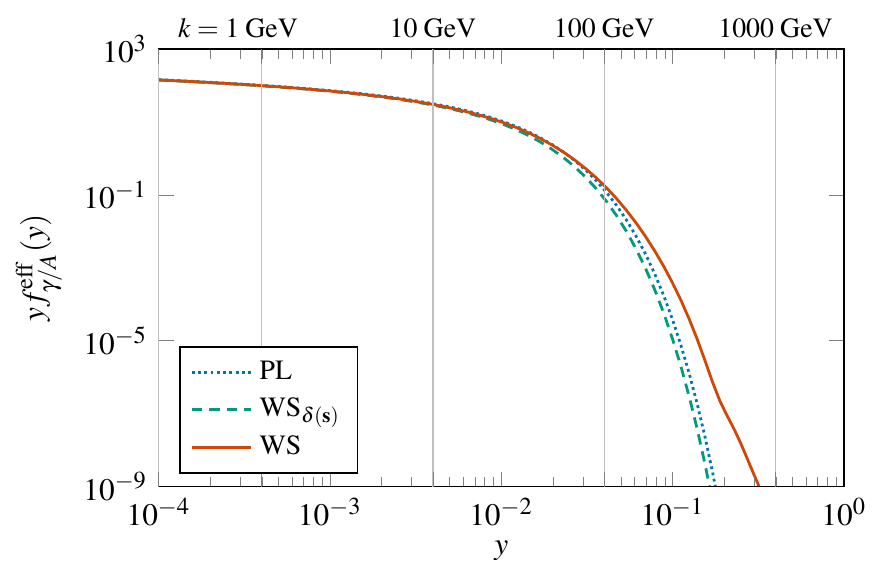}
    \includegraphics[width=8.cm]{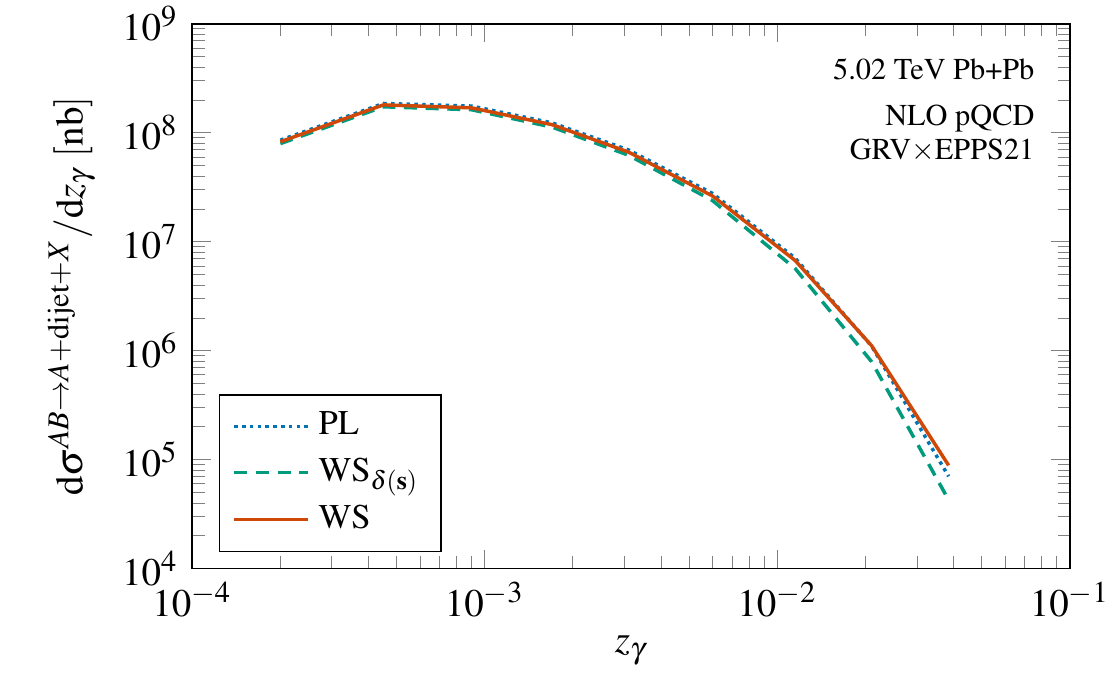}
    }
 \caption{(Left) The effective photon flux $f_{\gamma/A}^{\rm eff}$ in Pb-Pb UPCs at 5.02 TeV as a function of the photon 
momentum fraction $y$ (lower $x$-axis) and the photon energy $k=y\sqrt{s_{NN}}/2$ (upper $x$-axis), where the three curves correspond to the 
three approximations in Eq.~(\ref{eq:flux_effective}), see text for details.
 (Right) The NLO pQCD results for the cross section of inclusive dijet photoproduction in Pb-Pb UPCs at 5.02 TeV as a function of $z_{\gamma}$.
The three curves correspond to the three approximations for the effective photon flux shown on the left.
}
\label{fig:upc_dijets}
\end{figure}

The left panel of Fig.~\ref{fig:upc_dijets} (the figure is from~\cite{Eskola:2024fhf}) shows the effective photon flux $f_{\gamma/A}^{\rm eff}$ in Pb-Pb UPCs at $\sqrt{s_{NN}}=5.02$ TeV as a function of the photon 
momentum fraction $y$ (lower $x$-axis) and the photon energy $k=y\sqrt{s_{NN}}/2$ in the c.m.~frame (upper $x$-axis). The red solid curve corresponds to 
the calculation using Eq.~(\ref{eq:flux_effective}), where all the involved functions are calculated using the Wood-Saxon (WS) charge and nuclear density distributions; the green dashed curve corresponds to the $T_A(\vec{s}) \to \delta^2(\vec{s})$ approximation, which assumes the impact-parameter averaged nPDFs; and the blue dotted line is the PL photon flux.
One can see from the figure that while the two latter approximations lead to the essentially indistinguishable predictions, taking into account the 
impact-parameter dependence of nPDFs noticeably increases the effective photon flux for large $y$ and $k$.

The right panel of Fig.~\ref{fig:upc_dijets} presents the NLO pQCD results for the cross section of inclusive dijet photoproduction in Pb-Pb UPCs at 5.02 
TeV (\ref{eq:upc_dijets}) as a function of $z_{\gamma} \approx y x_{\gamma}$, the parton momentum fraction on the photon side in the resolved photon case or the photon momentum fraction in the direct photon case. 
The calculation uses the cuts of the ATLAS analysis~\cite{ATLAS:2024mvt}, the EPPS21 nPDFs, and the GRV photon PDFs;
the three 
curves correspond to the three approximations for the effective photon flux shown in the left panel of Fig.~\ref{fig:upc_dijets}.
One can see from the figure while they give very similar results for low $z_{\gamma}$, they begin to significantly differ for large $z_{\gamma}$. Indeed, large values of $z_{\gamma}$ correspond to large photon energies $k$ and, hence, to smaller than average, ``near-encounter`` impact parameters $\vec{b}$, where one is sensitive to the transverse plane geometry.  

In addition to the cuts on jet variables, the ATLAS analysis also used the 0nXn event selection criterion, requiring no neutrons detected in a zero degree calorimeter (ZDC)
on one side and at least one neutron on the other side. To take this condition into account, one needs to modify the effective photon flux 
$f_{\gamma/A}^{\rm eff}$ by adding to Eq.~(\ref{eq:flux_effective}) the factor of $\Gamma_{AA}^{\rm e.m.}(\vec{b})$, which gives the probability for not having the electromagnetic breakup of the photon-emitting nucleus. This reduces the total survival probability $\Gamma_{AA}^{\rm e.m.}(\vec{b})\Gamma_{AA}(\vec{b})$ and has a visible impact on the calculated dijet cross section. The latter can be quantified by the fraction 
$f_{\rm no\, BU}=0.2-0.6$, which corrects the cross section for the effect of breakup of the photon-emitting nucleus. This effect is quite 
significant and has been accounted in the data-to-theory comparison by ATLAS~\cite{ATLAS:2024mvt}.

\section{Inelastic $\gamma A$ scattering accompanied by forward neutrons in ZDCs}
\label{sec:zcd}

In addition to production of light and heavy vector mesons (Sec.~\ref{sec:Jpsi}) and dijets (Sec.~\ref{sec:dijets}) in heavy ion UPCs, 
it is important to consider other processes (see Sec.~\ref{sec:summary}) providing new constraints on the nuclear parton structure and small-$x$ dynamics of
the strong interaction. One such an example is inelastic $\gamma A$ scattering leading to production of forward neutrons from nuclear breakup, which can be detected by zero degree calorimeters (ZDCs) at the LHC~\cite{Alvioli:2024cmd}. 
The number of produced neutrons is correlated with the number of inelastic photon-nucleon interactions (wounded nucleons), which in turn encode information
on nuclear shadowing, including its impact parameter dependence.  

Below we outline the main idea. Generalizing the Abramovski-Gribov-Kancheli (AGK) cutting rules~\cite{Bertocchi:1976bq} to photon-induced processes mediated by hadronic components of the photon (the resolved photon contribution),
the photon-nucleus inelastic cross section $\sigma_{\rm inel}^{\gamma A}$ can be presented in the following form
\begin{equation}
\sigma_{\rm inel}^{\gamma A}=\sum_{\nu=1}^{A} \sigma_{\nu}=\sum_{\nu=1}^{A} \frac{A!}{(A-\nu)! \nu!} \int d^2 \vec{b} \int d\sigma
  P_{\gamma}(\sigma) (\sigma_{\rm inel} T_A(\vec{b}))^{\nu} (1-\sigma_{\rm inel} T_A(\vec{b}))^{A-\nu} \,,
\label{eq:sigma_inel}
\end{equation}
where $P_{\gamma}(\sigma)$ is the probability for hadronic fluctuations of the photon to interact with target nucleons with the cross section $\sigma$~\cite{Frankfurt:2022jns,Alvioli:2016gfo};
$T_A(\vec{b})=\int dz \rho_A(\vec{b},z)$ and $\rho_A(\vec{b},z)$ is the nuclear density depending on the transverse coordinate (impact parameter) $\vec{b}$ and
the longitudinal position $z$;
$\sigma_{\rm inel}=0.85 \,\sigma$ is an estimate for the inelastic cross section. 
The partial cross sections $\sigma_{\nu}$ correspond to the process,
where $\nu$ nucleons undergo inelastic scattering, while the remaining $A-\nu$ nucleons provide absorption. 
In the literature, one often uses the term ``wounded nucleons'' for $\nu$.

\begin{figure}[t!]
  \centerline{%
    \includegraphics[width=8cm]{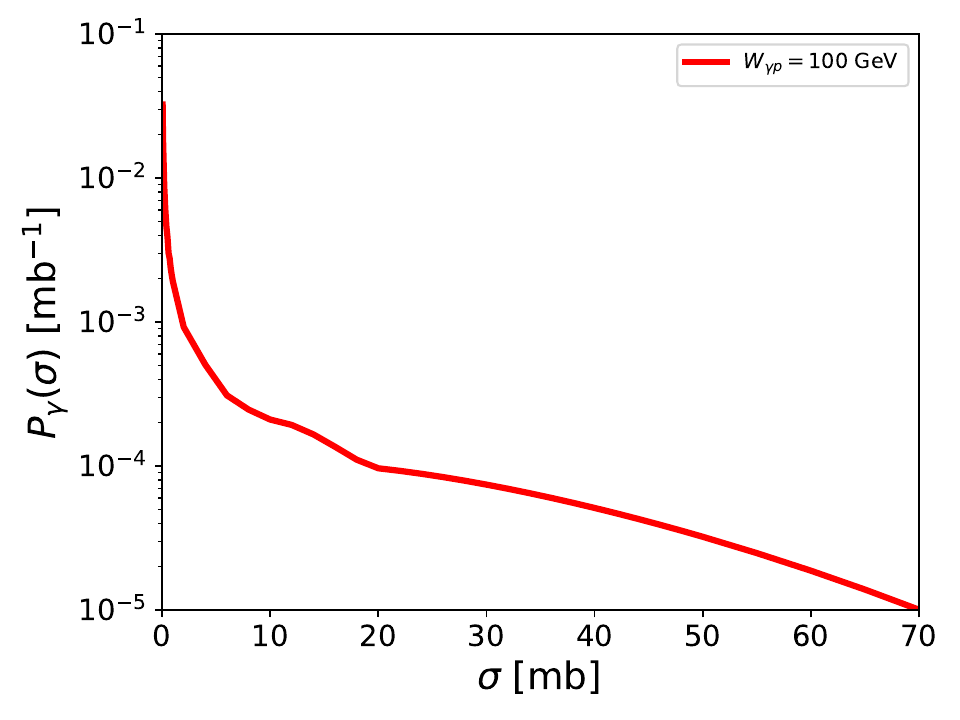}
    \includegraphics[width=8cm]{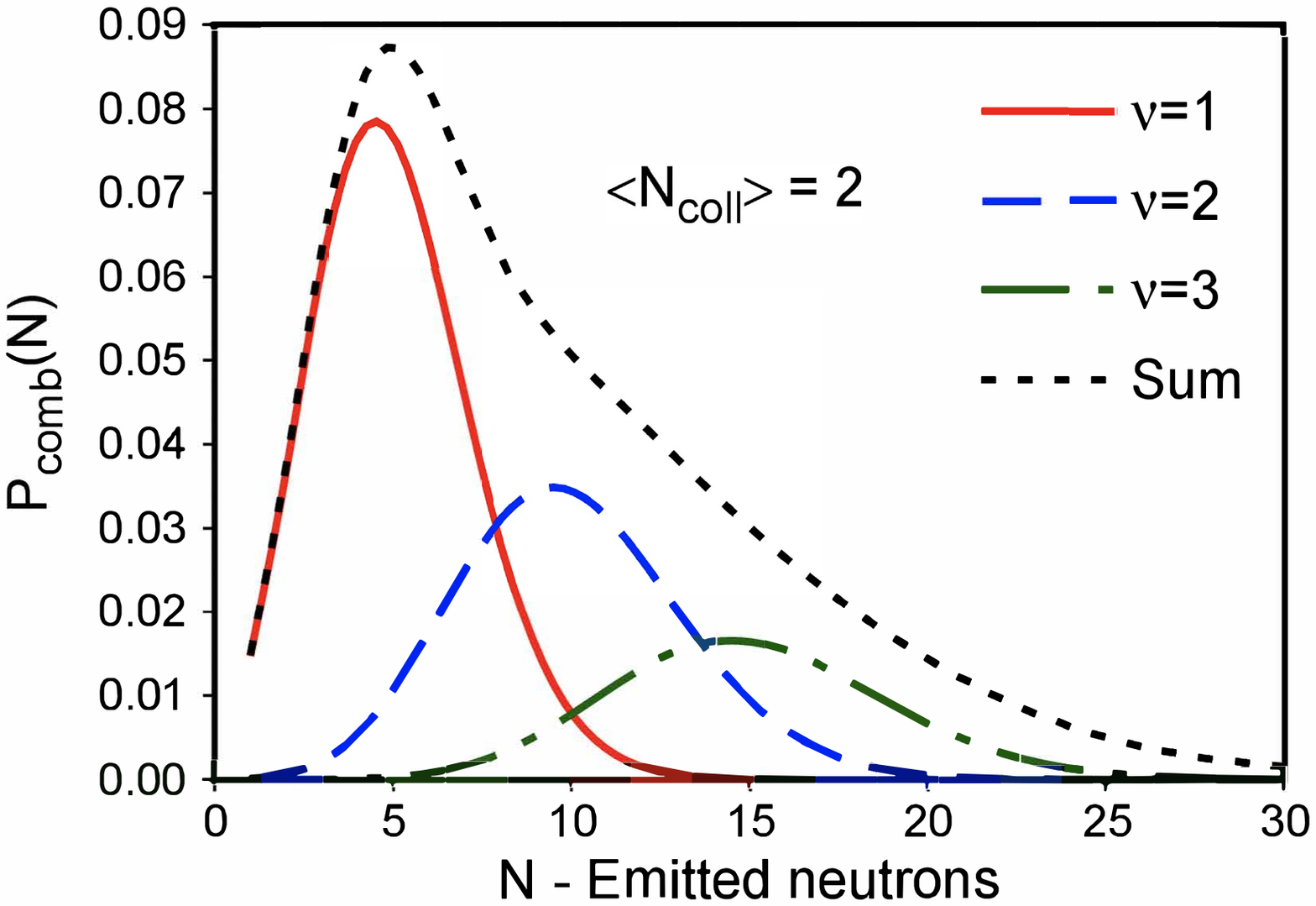}
    }
 \caption{(Left) The distribution $P_{\gamma}(\sigma)$ as a function of the cross section $\sigma$ at $W_{\gamma p}=100$ GeV.
 (Right) The probability of forward neutron emission $P_{\rm comb}(N)$~(\ref{eq:P_comb}) as a function of the number of emitted neutrons $N$ 
 at $\langle N_{\rm coll} \rangle=2$: the separate contributions of $\nu=1,2,3$ wounded nucleons and the full result.
}
\label{fig:Pgamma}
\end{figure}

The distribution $P_{\gamma}(\sigma)$ as a function of the cross section $\sigma$ at $W_{\gamma p}=100$ GeV is shown in the left panel of Fig.~\ref{fig:Pgamma}. It is constructed as a smooth interpolation
between the small-$\sigma$ region, where $P_{\gamma}(\sigma)$ can be expressed in terms of the quark-antiquark component of the photon wave function and the 
color dipole cross section leading to $P_{\gamma}(\sigma) \propto 1/\sigma$, and the region of large $\sigma$, where $P_{\gamma}(\sigma)$ can be modeled using
the hadronic fluctuations of the $\rho$ meson (the pion). Since the dependence of $P_{\gamma}(\sigma)$ on $W_{\gamma p}$ is weak, the distribution in Fig.~\ref{fig:Pgamma} is applicable to a wide range of energies probed in heavy ion UPCs at the LHC.

Using Eq.~(\ref{eq:sigma_inel}), the average number of inelastic photon-nucleon collisions (so-called wounded nucleons) $\langle N_{\rm coll} \rangle$
can be expressed as follows
\begin{equation}
\langle N_{\rm coll} \rangle \equiv \langle \nu \rangle = \sum_{\nu=1}^A P(\nu) \nu =\frac{\sum_{\nu=1}^A \nu \sigma_{\nu}}{\sum_{\nu=1}^A \sigma_{\nu}}= \frac{A \sigma_{\rm inel}^{\gamma N}}{\sigma_{\rm inel}^{\gamma A}} \approx \frac{1}{S_{Pb}(x)} \lesssim 2 \,,
\label{eq:N_coll}
\end{equation}
where $P(\nu)=\sigma_{\nu}/\sum_{\nu=1}^A \sigma_{\nu}$ is the normalized distribution of the number of wounded nucleons, and
$\sigma_{\rm inel}^{\gamma N}=\int d\sigma P_{\gamma}(\sigma) \sigma_{\rm inel}$ is the photon-nucleon inelastic cross section.
Equation~(\ref{eq:N_coll}) presents an example of the AGK cancellation due to unitarity of the underlying Gribov-Glauber model. It  allows one to relate 
$\langle N_{\rm coll} \rangle$ to the nuclear shadowing effect in $\sigma_{\rm inel}^{\gamma A}$, which can be approximated by the nuclear suppression factor
introduced by Eq.~(\ref{eq:S_Pb}) in Sec.~\ref{sec:Jpsi}.

It was experimentally established in muon-lead DIS at Fermilab that each muon-nucleon inelastic scattering releases on average 5 neutrons~\cite{E665:1995utr}.
Assuming that it happens independently for all $\nu$ and that the probability to emit $N$ neutrons is given by the Poisson distribution with the expectation
$\nu \langle M_n \rangle$, where $\langle M_n \rangle=5$, one obtains the following simple model for the combined probability of neutron emission
$P_{\rm comb}(N)$ in inelastic $\gamma A$ scattering~\cite{Alvioli:2024cmd}
\begin{equation}
P_{\rm comb}(N)=\sum_{\nu=1}^A P(\nu) \frac{(\nu \langle M_n \rangle)^{N}  e^{-\nu \langle M_n \rangle}}{N!} \,.
\label{eq:P_comb}
\end{equation}

The right panel of Fig.~\ref{fig:Pgamma} (the figure is from~\cite{Alvioli:2024cmd}) presents the probability of forward neutron emission as a function of the number of emitted neutrons $N$:
the curves present the separate contributions of $\nu=1,2,3$ wounded nucleons and the full result, see Eq.~(\ref{eq:P_comb}), and correspond to
the typical average number of wounded nucleons $\langle N_{\rm coll} \rangle=2$, see Eq.~(\ref{eq:N_coll}).
One can see from the figure that since the peaks corresponding to different $\nu$ are sufficiently separated, the measurement of $P_{\rm comb}(N)$ at different
$\langle N_{\rm coll} \rangle$ allows one in principle to reconstruct the distribution $P(\nu)$. The latter encodes complementary information on nuclear 
shadowing, including its impact parameter dependence, and allows one to study it ``one nucleon at a time''.

\section{Conclusions}
\label{sec:summary}

There are a continuing interest and a strong theoretical support for UPC studies at the LHC and RHIC.
On the one hand, the measurements of photon-induced processes is an essential part of the heavy ion program at the LHC, where they complement
$pA$ scattering. On the other hand, UPCs serve as a testing ground for $eA$ collisions at the planned EIC.
Thus, UPCs provide unique opportunities to obtain new information on the proton and nucleus partonic structure and the QCD dynamics at small $x$.

The measurements of coherent $J/\psi$ photoproduction in Pb-Pb UPCs at the LHC have revealed a strong suppression of the nuclear cross section compared to its impulse
approximation estimate. While the comprehensive theoretical description of these data is challenging both in the frameworks of collinear pQCD and the color dipole model, it can be interpreted as the first direct  evidence of the large leading twist gluon nuclear shadowing predicted by LTA and nuclear PDFs.
Note, however, that now there is a sizable discrepancy between the Run 2 ALICE and Run 3 ATLAS data at forward rapidities.
Thus, to clarify the situation, it is important to analyze all the available data on  $J/\psi$ production in Pb-Pb UPCs, including the $t$ dependence
of $d\sigma^{\gamma Pb \to J/\psi Pb}/dt$ as well as incoherent $J/\psi$ production. 
On the experimental side, for the unambiguous interpretation of the Pb-Pb data in terms of the nuclear suppression factor, it is essential to
have measurements of the proton baseline cross section, which can be achieved, e.g., using coherent $J/\psi$ production in $p$Pb UPCs~\cite{Guzey:2013taa}.

An important cross-check of models for nuclear shadowing is provided by the recent data on coherent $\Upsilon$ photoproduction in Pb-Pb UPCs, which allow one study the momentum fraction $x$ and the resolution scale $Q^2$ dependence of the gluon nuclear shadowing in the collinear framework.

Inclusive dijet photoproduction in Pb-Pb UPCs at the LHC provides a complementary probe of nuclear PDFs at small $x$ down to 
$x_A=0.002$ for $\sqrt{Q^2} \geq 35$ GeV and can in principle reduce the current small-$x_A$ uncertainties of the gluon
distribution~\cite{Guzey:2019kik}. The precision QCD analysis of these data requires taking into account the 
transverse plane geometry and the experimental neutron class condition as well as the correction for the contribution of the diffractive dijet photoproduction~\cite{Guzey:2020ehb}.

One continues to explore the full potential of UPCs by studying additional processes. One recent suggestion includes inelastic $\gamma A$ scattering
accompanied by forward neutrons detected in a ZDC, where the measurement of the neutron spectrum allows one in principle to reconstruct the distribution of wounded nucleons, which probes nuclear shadowing ``one nucleons at a time''.

The other new and promising process is inclusive $D^0$ production in Pb-Pb UPCs~\cite{CMS:2025jjx}, which provides additional, small $x$ and intermediate $Q^2$, constraints on the nPDFs~\cite{Cacciari:2025tgr} on the one hand and on the color glass condensate (CGC) framework~\cite{Gimeno-Estivill:2025rbw} as well as the unintegrated gluon distributions~\cite{Goncalves:2025wwt} on the other hand.

\section*{Acknowledgments}

The research of V.G.~was funded by the Academy of Finland project 330448, the Center of Excellence in Quark Matter
of the Academy of Finland (projects 346325 and 346326), and the European Research Council project ERC-2018-ADG-835105 YoctoLHC.


\end{document}